# UltraFast Optical Imaging using Multimode Fiber based Compressed Sensing and Photonic Time Stretch

GUOQING WANG, [1] CHAITANYA K MIDIDODDI, [1] FANGLIANG BAI, [2] STUART GIBSON, [2] LEI SU, [2] JINCHAO LIU, [2] CHAO WANG, [1,*]

[1]*School of Engineering and Digital Arts, University of Kent, Canterbury, United Kingdom CT2 7NT*
[2]*School of Physics Science, University of Kent, Canterbury, United Kingdom CT2 7NT*
*Corresponding author: author_three@uni-jena.de*

**An ultrafast single-pixel optical 2D imaging system using a single multimode fiber (MF) is proposed. The MF acted as the all-optical random pattern generator. Light with different wavelengths pass through a single MF will generator all-optical random speckle patterns, which have a low correlation of 0.074 with 0.1nm wavelength step from 1518.0nm to 1567.9nm. The all-optical random speckle patterns are perfect for compressive sensing (CS) imaging with the advantage of low cost in comparison with the conventional expensive pseudorandom binary sequence (PRBS). Besides, with the employment of photonic time stretch (PTS), light of different wavelengths will go through a single capsuled MF in time serial within a short pulse time, which makes ultrafast single-pixel all-optical CS imaging possible. In our work, the all-optical random speckle patterns are analyzed and used to perform CS imaging in our proposed system and the results shows a single-pixel photo-detector can be employed in CS imaging system and a 27 × 27 pixels image is reconstructed within 500 measurements. In our proposed imaging system, the fast Fourier transform (FFT) spatial resolution, which is a combination of multiple Gaussians, is analyzed. Considering 4 optical speckle patterns, the FFT spatial resolution is 50 × 50 pixels. This resolution limit has been obtained by removing the central low frequency components and observing the significant spectral power along all the radial directions.**

## 1. INTRODUCTION

In Single-pixel imaging, which overcomes the bottleneck of direct sight of the object, draws extensive attention due to its advantage of enabling high quality images acquiring with a single point detector [1-5]. However, owing to the usage of spatial light modulator (SLM) or digital micromirror devices (DMD) [1], the speed of single-pixel imaging systems is limited to tens of KHz.

To enhance the speed of the single-pixel imaging systems, several of methods are proposed, one of the most popular methods is using photonic time stretch (PTS) technique [6], also named dispersive Fourier transform [7, 8] and frequency-to-time mapping [9], which has been implemented in OCT [10, 11], ultrafast optical imaging systems [12-15], ultrafast waveform measurement [16, 17], and analog-to-digital technology [18, 19]. PTS can map the broadband optical spectrum of an ultrashort optical pulse into temporal waveform via the employment of group velocity dispersion. Hence, the temporal resolution of the optical system can be greatly enhanced and ultrafast optical measurement system is made possible via using a high-speed single-pixel photo-detector (PD).

However, the utilization of PTS leads to big data problem, that hundreds of gigabits even terabits of data needs to be processed in one second. Thus, big data compression technique is desired. Compressive sensing (CS) is a promising technique for big data compression and it is based on the theory that the sparse information in a transformation domain can be recovered from a reduced number of measurements, which conquer the traditional Nyquist-Shannon sampling theorem [20, 21]. Hence, it got pervasive attention in radio frequency (RF) signal detection [22, 23], optical coherence tomography (OCT) [10, 11, 24, 25] and CS-based imaging [12, 13].

To achieve big data compression using CS technique, pseudorandom binary sequence (PRBS) is widely employed to generate random serials [10, 11], which have the drawback of high cost and narrow bandwidth owing to the employment of expensive arbitrary waveform generator (AWG). Also, the speed of CS imaging is limited by multiple measurements to obtain one frame. As a result, ultrahigh-speed low-cost random sequential single-pixel imaging is required. In [2], a special TiO2-coated fiber tip, which could induce wavelength-dependent random speckle patterns, is used to eliminate the usage of expensive PRBS. However, the imaging speed of this methods is limited to the tunable laser speed and its iteration. Thus, another high-speed low-cost random sequential method is presented using multimode fiber (MF) [26]. The MF is used for CS thanks to MF causing randomization of phase, polarization and optical speckle pattern distribution when light travels through it. MF has bigger core, higher capacity and larger numerical aperture, which attracts extensively resurging attention in imaging [27-33] and communication [34, 35] applications.

Our proposed single-pixel imaging system, which is named UltraFast Optical Imaging system based MF (UFO-IMF), with combination of CS and PTS, has the unprecedented CS imaging speed under same laser source condition. The frame rate of our imaging system is equal to the repetition rate of the pulsed laser. Compared to traditional all-optical random pattern generator that using spatial light modulator (SLM) or digital micromirror devices (DMD), which generates hundreds or tens of thousands of random speckle patterns in one second, ultrashort optical pulse with broadband spectrum that passes through a single capsuled MF based PTS could introduce an all-optical random speckle pattern generation speed of tens of GHz, which is around 6 to 7 orders higher. When in comparison with other PRBS-based high-cost CS and PTS imaging system, such as OCT [10, 11, 24, 25] and serial time-encoded amplified microscopy (STEAM) [12-15], which need hundreds of times iteration of a single ultrashort optical pulse to reconstruct one frame, the low-cost frame rate of UFO-IMF is hundreds of times faster thanks to its fundamental nature.

In this work, we present our conceptual UFO-IMF with the employment of CS and PTS. Via the usage of MF, our proposed system has the benefit of ultrafast all-optical random speckle pattern generation speed with low cost and wide bandwidth in comparison with traditional high cost and narrow bandwidth PRBS generator and low speed SLM / DMD, also it has the advantage of inherent 2D imaging reconstruction in nature. To gain a better understanding of our proposed system, a combination of analysis with computation and experiment is presented. The theory of MF based PTS is presented. The all-optical random speckle patterns generated in a single capsuled MF via light of different wavelengths travelling through it in time serial are experimentally analyzed and demonstrated. An imaging compression ratio from 35.11% to 68.59% is shown in computation. Also CCD imaging at same pixel size is computed for comparison. The fast Fourier transform (FFT) spatial resolution is analyzed according to the all-optical random speckle patterns and the result in our system is 50 × 50 pixels.

## 2. PRINCIPLES

The all-optical random speckle patterns are generated via the light of different wavelengths that travels in a capsuled MF. By the implementation of PTS, ultrashort optical pulse of broadband spectrum can be stretched in temporal domain. Each single wavelength will generate its own random and repeatable optical speckle pattern in time sequence. Thus all-optical random speckle patterns can be generated once the stretched pulses go through the MF. The process diagram is shown in Fig. 1. At each time, only one certain optical speckle pattern at one wavelength is generated (the mode dispersion can be neglected if selectively choose short length MF and control the incident light when propagating through MF). However, the time difference between each wavelength is extremely short, which could be around 0.1ns. The traditional imaging method for photon-detection and data acquisition in such ultrafast speed needs bandwidth of trillions of Hz and sampling speed of trillions of samples per seconds. In our case, CS technique is employed and a 10GHz bandwidth single-pixel PD can be used to reconstruct the image.

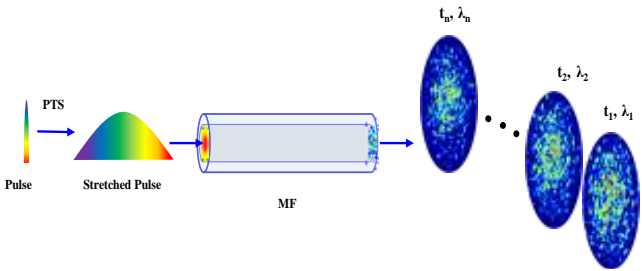

Fig. 1. The process diagram of pulse stretch and all-optical random speckle pattern generation via the combination of PTS and MF. PTS: photonics time stretch; MF: multimode fiber.

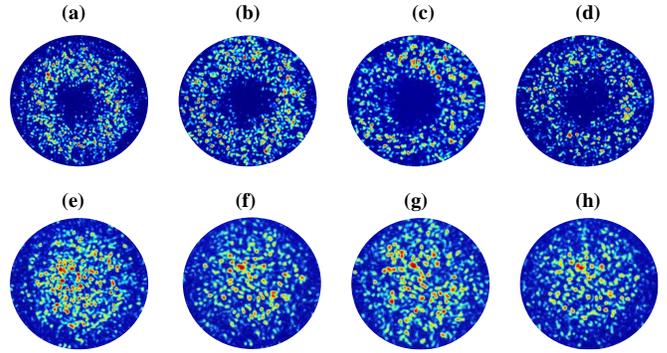

Fig. 2. The weak mode coupling of MF at 1520nm (a), 1530nm (b), 1540nm (c), 1550nm (d); strong mode coupling of MF at 1520nm (e), 1530nm (f), 1540nm (g), 1550nm (h).

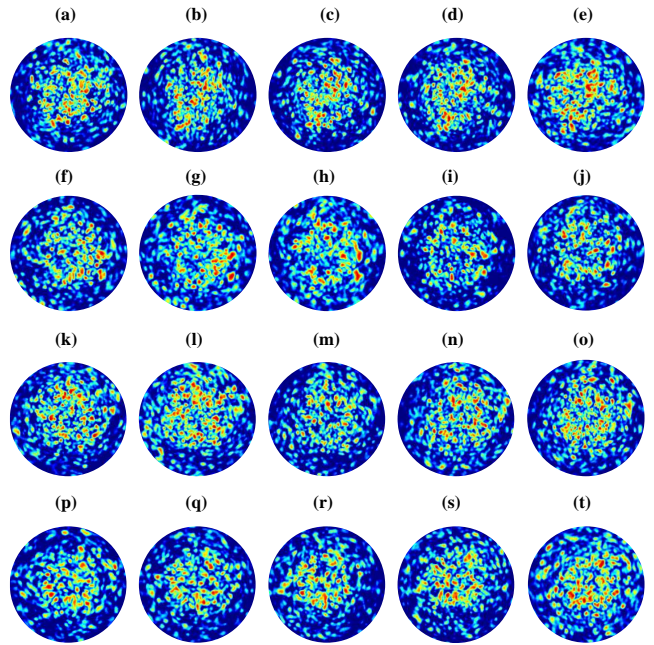

Fig. 3. All-optical random speckle patterns at (a) – (e) 1520.0 – 1520.4nm; (f) – (j) 1530.0 – 1530.4nm; (k) – (o) 1540.0 – 1540.4nm; (p) – (t) 1550.0 – 1550.4nm, with 0.1 nm step.

The nonlinear wave equation [36] depicts the transmission of 2D speckle field in MF,

$$i\frac{\partial E}{\partial z} + \frac{1}{2k_0 n_0}\left[\frac{\partial^2}{\partial x^2} + \frac{\partial^2}{\partial y^2}\right]E + k_0 n_2 |E|^2 E = 0 \quad (1)$$

where n2 is the parameter that describes the speckles will be focus ( n2 > 0 ) or defocus ( n2 < 0 ) after transmitted in the MF [37]. In our experiment the results show n2 < 0, although this could reduce the spatial resolution of our imaging system, the randomization and spatial intensity distribution, which determines the success of our experiment, performs much better. Besides, the weak mode coupling and strong mode coupling is investigated as this could affect the experimental result [38]. As shown in Fig. 2 (a) to (d), weak mode coupling from wavelength 1520 nm to 1550nm illustrates the unevenly random spatial intensity distribution, which is not qualified for imaging; Fig. 2 (e) to (h) reveal the strong mode coupling from wavelength 1520 nm to

1550nm and their random spatial intensity distribution, which scattered to the whole imaging area and could be employed for imaging application. Also the strong mode coupling is more energy efficient for its low light coupling loss. There is a trade-off between the size of the all-optical speckles and their randomization. To increase the randomization of the all-optical speckles, the spatial resolution, which determined by the size of the all-optical speckles, is sacrificed. Part of the all-optical speckle patterns used in our experiment is shown in Fig. 3, all-optical random speckle patterns at four wavelength bands are shown and each band has 5 wavelength-dependent random speckle patterns with 0.1nm wavelength step.

In our CS computation, a target 2D image $I_{M \times M}$ is mixed with N random patterns, each $R_{M \times M}$ corresponding to a specific wavelength. N also corresponds to the number of measurements. By vectoring the image into 1D ($I_{MM \times 1}$) and N random patterns ($R_{MM \times N}$), the CS system can be simplified into 1D model. The image $I_{MM \times 1}$ assumed to be sparse in discreet Fourier transformation (DFT) domain, the measurement vector $y_{N \times 1}$ can be described as the dot product between the image and random patterns, $y_{N \times 1} = R_{N \times MM} \cdot I_{MM \times 1}$. Image $I_{MM \times 1}$ can be represented in transformation domain $\varphi_{MM \times MM}$ as $s_{MM \times 1} = \varphi_{MM \times MM} \cdot I_{MM \times 1}$, where $s_{MM \times 1}$ denotes the transformation domain representation. Hence the equations can be summarized as $y_{N \times 1} = R_{N \times MM} \cdot \varphi^{-1} s_{MM \times 1} = \theta_{N \times MM} \cdot s_{MM \times 1}$, where $\theta_{N \times MM} = R_{N \times MM} \cdot \varphi^{-1}{}_{MM \times MM}$.

In the imaging reconstruction region, the measurement vector $y_{N \times 1}$ and $R_{N \times MM}$ are given as inputs and the transformation domain $\varphi_{MM \times MM}$ is known, Hence $s_{MM \times 1}$ can be reconstructed using total variation (TV) minimization algorithm, $(TV_1) \min TV(s)$ subject to $\theta s = y$ [39]. From the retrieved transformation representation output $\hat{s}$, the image can be reconstructed as, $\hat{I}_{MM \times 1} = \varphi_{MM \times MM} \hat{s}_{MM \times 1}$ and can be devectorized to get the 2D image $\hat{I}_{M \times M}$ [40].

## 3. SIMULATION, EXPERIMENT AND RESULT

The schematic of UFO-IMF based CS and PTS is shown in Fig. 4. The capsuled MF in this setup is performed as a low-cost, ultrafast and passive all-optical random speckle generator and the capsulation is performed to minimize the effect of the environment, which can fatally affect the stability of the all-optical random speckle patterns. A mode-locked laser (MLL) with a repetition rate of 20MHz is utilized to generate the ultrashort pulses with broadband spectrum. The ultrashort pulses go through the dispersive compensating fiber (DCF, with a whole dispersion of 1ns/nm) to perform PTS, namely, dispersive Fourier transform. Therefore, wavelength-to-time one-to-one mapping is achieved. The time stretched pulses are emitted into open space via a collimator (Col) with a numerical aperture (NA) of 0.49. Then the pulses are coupled into capsuled MF (with a length of 2m and the diameter of its core is 200μm, 0.39NA) via pass through objective lens (×10, 0.40NA), which induces more modes that could improve the randomization of the optical speckles. When each of the time stretched broadband spectrum pulses propagates through the capsuled MF, each

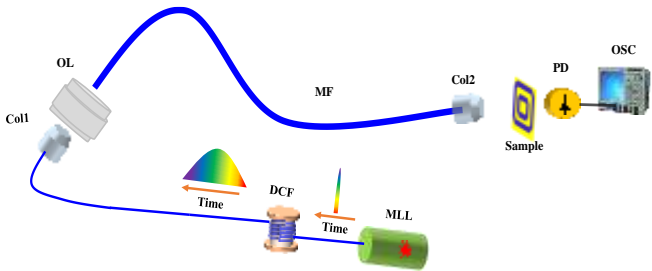

Fig. 4. Schematic of UFO-IMF based CS and PTS. UFO-IMF: ultrafast optical 2D imaging using multimode fiber; CS: compressed sensing; PTS: photonic time stretch; MLL: mode-locked laser; DCF: dispersive compensating fiber; Col: collimator; OL: objective lens; MF: multimode fiber; PD: photo-detector; OSC: oscilloscope.

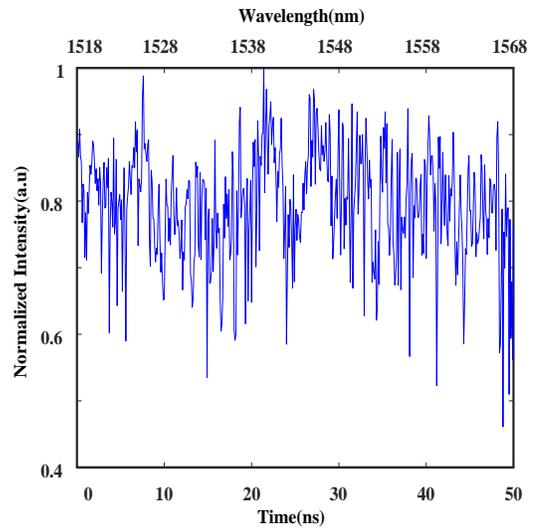

Fig. 5. The computed signal received by PD in temporal domain (500 all-optical random speckle patterns).

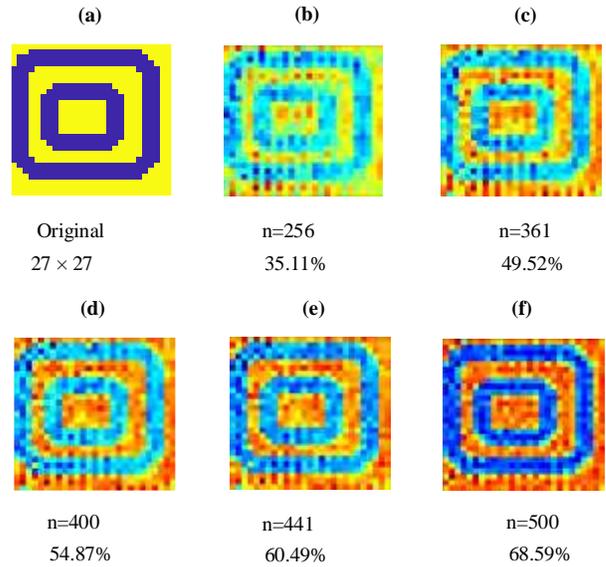

Fig. 6 (a) An original target image with pixels size 27 × 27; (b) to (f), reconstructed images used different number of all-optical random optical speckle patterns (different compression ratios from 35.11% to 68.59%); (g) to (k), CCD images downsized from (a) with corresponding pixels size as (b) to (f).

wavelength in a pulse generates its own random, stable and repeatable 2D all-optical speckle pattern at its own time point. After the all-optical random speckle patterns in time serial pass through the target image, a single-pixel PD with properly chose bandwidth is used to receive the data. The final data are collected by an oscilloscope (OSC) with a sampling rate the same as the bandwidth of the PD numerically.

The computed data received by PD is illustrated in. 5. A PD with a bandwidth of 10GHz is employed. The MLL has a wavelength range of 1518 nm to 1568nm in simulation and the DCF has a dispersion of 1ns/nm. Every 0.1ns is a measurement in our CS technique.

An original customer-designed 27 × 27 pixels image (shown in Fig. 6 (a)) is employed in our setup as the target image to perform CS technique. Fig. 6 (b) to (f) show the reconstructed images with different number of all-optical random speckle patterns that used in CS technique, which is from 256 to 500 (with compression ratios from 35.11% to 68.59%). And from the result we could get the conclusion that the CS technique in our proposed imaging system can work as expected to fulfill the low-cost ultrafast imaging purpose with distingu-

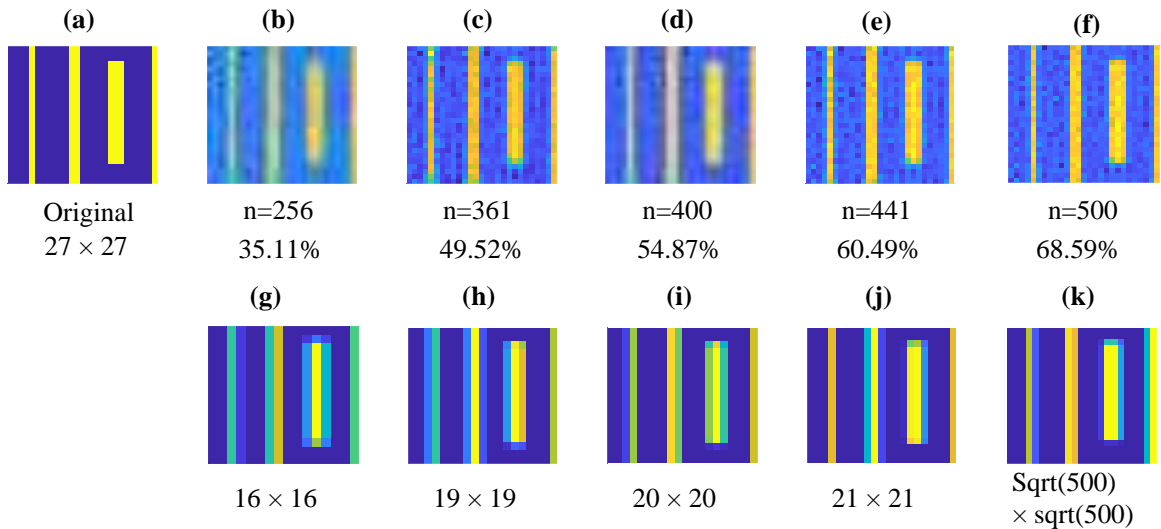

Fig. 7 (a) An original target image with pixels size 27 × 27; (b) to (f), reconstructed images used different number of all-optical random speckle patterns (different compression ratios from 35.11% to 68.59%).

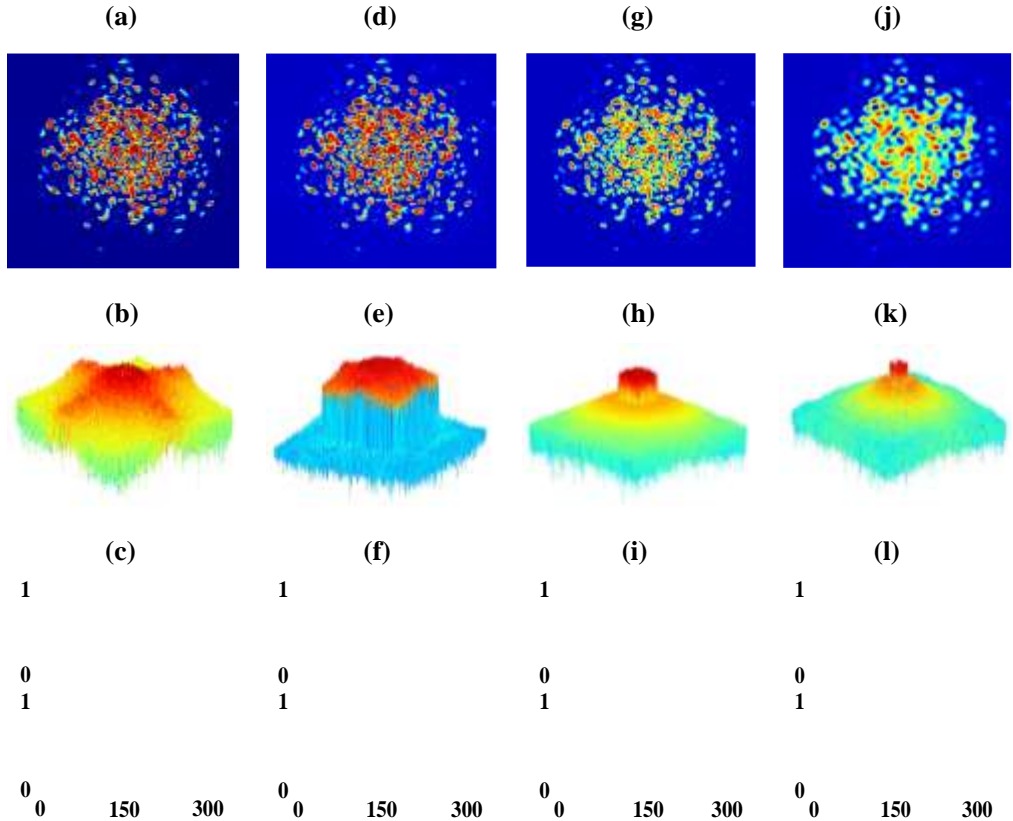

Fig. 8. (a) Pattern captured by beam profiler with 540 × 540 pixel resolution; (b) 2D spectral domain representation after removing the low frequency components; (c) Superimposed spectral domain representation of individual rows of the image followed by superimposed representations of the columns of the image shown in (a); (d) The random pattern after removing the high spatial frequency low spectral power components; (e) Corresponding 2D spectral FFT of (d); (f) superimposed representations after removal of high frequency components; (g) The random pattern after removing the high spatial frequency low spectral power components; (h) Corresponding 2D spectral FFT of (g); (i) superimposed representations after removal of high frequency components; (j) The random pattern after removing the high spatial frequency low spectral power components; (k) Corresponding 2D spectral FFT of (j); (l) superimposed representations after removal of high frequency components.

ishable resolution, and the more speckle patterns that used in CS technique, the more details can be identified in the reconstructed image.

An original four-line-target image with pixels size 27 × 27, which is shown in Fig. 7 (a), is utilized to exemplify the resolution effect of CS imaging. Reconstructed images with different compression ratios / numbers of speckle patterns used from 35.11% / 256 to 68.59% / 500 are depicted from Fig. 7 (b) to (f), respectively. The result shows the CS technique in our proposed imaging system matches quite well with the target image. To compare the imaging effect of our CS technique, different CCDs with same pixels size as Fig. 7 (b) to (f) are employed to receive the target image of Fig. 7 (a), and the images are shown in Fig. 7 (g) to (k). Compared to CCD imaging with same pixels size respectively, the CS imaging can give better resolution in the target image although with the sacrifice of signal-to-noise ratio (SNR), such as in comparison with Fig. 7 (h) and (k), (c) and (f) show more accurate details in the edge areas, especially notable with the first and last narrow lines.

## 4. DISCUSSION

### A. Spatial resolution

The fast Fourier transform (FFT) spatial resolution of our system is analyzed based on the random speckle patterns we used in the CS process. Given a range of all-optical random speckle patterns generated in the MF, the resolution is estimated to be 50 × 50 pixels.

Though the pixel resolution is 540 × 540, the random pattern cannot resolve the image of that size as the spatial resolution is limited by the random distance among the speckles. Here we demonstrate the actual random pattern captured by the beam profiler along with spectral profile of the random pattern followed by several spatial low pass filters to remove the high frequency components to see any significant difference in the pattern. As observed, the actual Fig captured by beam profiler with 540 × 540 pixel resolution is shown in Fig. 8 (a). The corresponding 2D spectral domain representation is shown in Fig. 8 (b), after removing the dominant low frequency component with a low pass filter and actual spectral power variation is shown in Fig. 8 (c), where the spectral power significantly decreases on higher spatial frequencies.

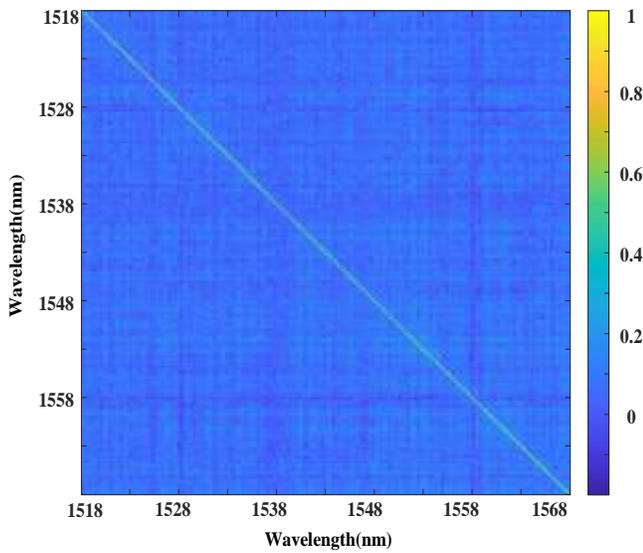

Fig. 9. Correlation of 500 optical speckle patterns from 1518.0 to 1567.9 nm with 0.1 nm step.

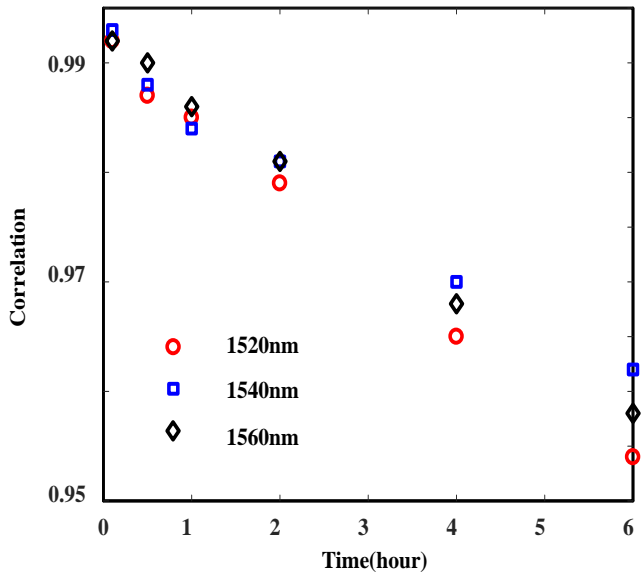

Fig. 10. The temporal stability of optical speckle patterns.

The higher spatial frequencies >150 are now suppressed using a low pass filter and the result is shown in Fig. 8 (f). The corresponding random pattern is shown in Fig. 8 (d), and the corresponding 2D FFT representation is shown in Fig. 8 (e).

Similarly the superimposed spectral domain representations of the pattern after removing the higher spatial frequencies >50 are shown in Fig. 8 (i), corresponding random pattern is shown in Fig. 8 (g), and 2D FFT representation is shown in Fig. 8 (h). The procedure is repeated for spatial frequencies >25 and the results are represented in Fig. 8 (l), (j), (k), respectively. As observed from the patterns with original Fig. 8 (a) and suppressed high frequency random pattern in Fig. 8 (j), some of the high frequency features have been lost marginally and this can be considered Nyquist frequency limit for the system. Hence any 50 × 50 image can be successfully reconstructed by the CS system.

### B. Speckle correlation

One major principle must be satisfied in CS technique is that the random speckle patterns should be stable, repeatable, while at the same time all the other patterns should be completely uncorrelated. The calibration process of all-optical random speckle patterns is performed via tuning wavelength using a tunable laser. The 20 wavelength-dependent optical speckle patterns generated by MF is shown in Fig. 3, from wavelength 1520.0 nm (a) to 1550.4 nm (t), respectively. The correlation of 500 stable and repeatable wavelength-dependent all-optical random speckle patterns are analyzed and the correlation result is shown in Fig. 9. The wavelength range from tunable laser is 1518.0 to 1567.9 nm with 0.1nm tuning step. The ideal correlation between every other two wavelength-dependent random speckle patterns should be as close to 0 and the self-correlation of one certain wavelength-dependent random speckle pattern should be 1. In Fig. 9, the average correlation value among every two different patterns in the 500 patterns is 0.074, and the self-correlation value of the 500 patterns is 1.

### C. Temporal speckle stability

The temporal stability of the all-optical random speckle patterns is analyzed via the correlation of previous and latter pattern within same wavelength at different captured time. The result is described in Fig. 10, three wavelengths of 1520 nm, 1540 nm and 1560 nm is used to exemplify the stability of all-optical random speckle patterns within 6 hours. The results show within one hour, all the patterns have correlation values of more than 0.985, although the value decreased as time increasing, the value is still above 0.95 within 6 hours. The calibration time of all-optical random speckle patterns is one pattern per second, so the whole calibration time of 500 patterns we used is 500 seconds. The frame speed in our proposed imaging system is 20MHz. Thus to make our proposal work, the all-optical random speckle patterns should be stable at least excess the time of calibration and imaging.

### D. Mode dispersion

The mode dispersion [41-43] is also considered in our imaging system owning to the mode dispersion can temporally stretch the pulse when we use a single pixel PD to receive the final data. However, when carefully choose the length, core diameter, NA of the MF and the coupling angle between the OL and MF, the mode dispersion can be decreased, such as in our situation, a step-graded MF with a length of 2m, core diameter 200μm, and 0.39 NA is employed. Fig. 11 (a) shows an original modulated single wavelength (1530nm) signal that propagating through MF with a half width at full maximum (HWFM) of 1.57ns ± 0.04ns with a repetition of every 20ns, and Fig. 11 (b) shows the output signal that stretched via mode dispersion has a HWFM of 1.60 ± 0.05ns. In practical this effect might affect the experiment as there may have part of information overlapping between the two consecutive measurements (0.1ns division), while in [44], a proposal that could solve 98% of overlapping between the two consecutive measurements and in theory this effect can be eliminated.

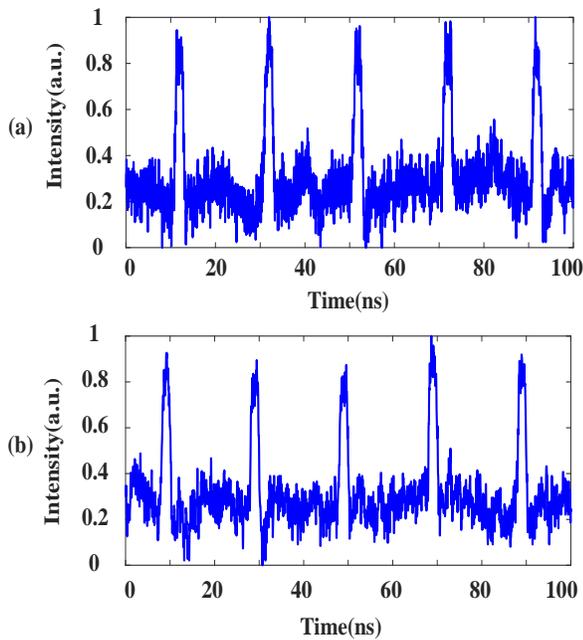

Fig. 11 (a) Incident signal with a HWFM of 1.57ns ± 0.04ns; (b) output signal after MF with a HWFM of 1.60ns ± 0.05ns. HWFM: half width at full maximum; MF: multimode fiber.

## 5. CONCLUSSION

In summary, UFO-IMF based CS and PTS is presented conceptually with demonstration. MF is regarded as an ultrafast all-optical random speckle pattern generator when combined with CS and PTS. This new conceptual all-optical random 2D speckle pattern generator has the advantage of low cost and large bandwidth. In comparison with traditional expensive and bandwidth-limited PRBS generator, our proposal is a perfect CS imaging method with low cost, ultrafast speed and inherently 2D CS imaging. For the other CS imaging systems that employed with SLM or DMDs, which have a maximum bandwidth tens of KHz due to the usage of electrical devices, our proposed imaging system has a bandwidth as high as 10 GHz, which is around 6 to 7 orders higher. Beside, by the exemplification of PTS, our CS 2D imaging method can have a frame time the same as one pulse time, while the other CS-based imaging method, such as OCT or STEAM, need to repeat hundreds of times to get one frame. Thus our presented imaging system has an even higher bandwidth compared to the other conventional CS-based imaging systems. To present the state-of-art of our imaging system, a combination of demonstration with computation and experiment is presented. The imaging compression ratio from 35.11% to 68.59% is illustrated in CS approach and CCDs with same down-sized pixels are used to detect the target imaging to compare the effect of the CS approach. Also the FFT spatial resolution is analyzed according to the all-optical random speckle patterns and the result in our system is 50 × 50 pixels. The calibration of all-optical random speckle patterns is demonstrated in experiment. Besides, other elements that could affect UFO-IMF, which includes temporal stability of random optical pattern and mode dispersion of MF, are investigated in our proposed system.

**Funding**. …

**Acknowledgment**. …